\documentclass{ws-procs975x65}

\begin{document}

\title{Parity violation in $pp$ scattering and vector-meson
weak-coupling constants}

\author{C.~H. Hyun} 

\address{Institute of Basic Science, Sungkyunkwan University,
Suwon 440-746, Korea \\
E-mail: hch@meson.skku.ac.kr}

\author{C.-P. Liu}

\address{KVI, Zernikelaan 25, Groningen 9747 AA, The Netherlands \\
E-mail: C.P.Liu@KVI.nl}

\author{B. Desplanques}

\address{Laboratoire de Physique Subatomique et de Cosmologie, \\
(UMR CNRS/IN2P3-UJF-INPG), F-38026, Grenoble Cedex, France \\
E-mail: desplanq@lpsc.in2p3.fr}

\maketitle

\abstracts{
We calculate the parity-nonconserving longitudinal asymmetry 
in the elastic $\vec{p} p$ scattering at the energies where 
experimental data are available. In addition to the standard
one-meson exchange weak potential, the variation of the strong-coupling 
constants and the non-standard effects such as form
factors and $2 \pi$-exchange description of the $\rho$-exchange
potential are taken into account. 
With the extra effects, we investigate
the compatibility of the experimental data and the presently-known 
range of the vector-meson weak-coupling constants. 
}

\section{Introduction}

The measurements of the parity-nonconserving
(PNC) asymmetry in $\vec{p}\, p$ scattering at both low energies
(13.6 and 45 MeV) and at high energy (221 MeV) have been
expected to determine the $\rho NN$ and $\omega NN$ coupling
constants of a one-meson exchange model of the PNC $NN$ force.
The analysis has been
recently performed by Carlson \textit{et al.}\cite{csbg-prc02},
and they obtain a positive $\omega NN$ PNC coupling.
Though their results do not disagree with the largest range 
provided by Desplanques, Donoghue and Holstein (DDH)\cite{ddh80}, 
updated calculations of the weak-coupling 
constants do not give much room for positive values of the $\omega NN$
weak coupling\cite{des-npa80,fcdh-prc91}.

With the DDH ``best-guess" values of PNC meson-nucleon couplings,
the contribution of PNC effects in $pp$ scattering is dominated by
the $\rho$-meson exchange. 
When the low-energy data are well reproduced, 
it gives a result suppressed roughly by a factor of 2 at high 
energy.
In this work, we concentrate on
the $\rho$-meson exchange with various effects such as strong-coupling
constants of the $\rho$ and $\omega$ mesons, form factor in the strong
or weak vertices\cite{ber98,km-npa} 
and the two-pion resonance nature of the $\rho$ meson\cite{cb-npb74}.
We investigate the role of these effects and whether they can give results
that satisfy all the low and high-energy data simultaneously with
a negative $\omega NN$ weak-coupling constant.
We illustrate the most important features of the work only, and
the details will be given elsewhere\cite{lhd-pncpp}.

\section{Extra effects}


\noindent{\bf Coupling constants}

The concerned PNC potential reads
\begin{eqnarray}
V_{\rho}(\bm{r}) & = & 
-\frac{g_{\rho NN}}{m_{N}}
\left[h_{\rho}^{0}\,\,\bm{\tau}_{1}\cdot\bm{\tau}_{2}
+ \frac{1}{2} h^1_\rho (\tau^z_1 + \tau^z_2) 
+ \frac{h_{\rho}^{2}}{2\sqrt{6}}\,\,
(3\tau_{1}^{z}\tau_{2}^{z}-\bm{\tau}_{1}\cdot\bm{\tau}_{2})
\right]\times\nonumber \\
 &  & \Big((\bm{\sigma}_{1}-\bm{\sigma}_{2})\cdot\{\bm{p},\, f_{\rho+}(r)\}-(\bm{\sigma}_{1}\times\bm{\sigma}_{2})\cdot\hat{r}f_{\rho-}(r)\Big).\label{eq:modf1}\end{eqnarray}
In the one-meson-exchange description, $f_{\rho +}$ is the usual
Yukawa function, ${\textrm{e}}^{-m_{\rho}r}/4\pi r$, and $f_{\rho-}(r)$
is its derivative with respect to $r$ (with the factor $(1+\kappa_V)$).
The relevant PNC coupling constants are 
$h^{pp}_\rho \equiv h^0_\rho + h^1_\rho + h^2_\rho/\sqrt{6}$ and
$h^{pp}_\omega \equiv h^0_\omega + h^1_\omega$ and their 
``best-guess" values are $-15.5$ and $-3.04$, respectively.
We consider the variations on the strong-coupling constants 
$g_{\rho NN}$ and $\kappa_V$ as well as $g_{\omega NN}$ and $\kappa_S$.
Our choices of the strong-coupling constants are summarized in 
Tab.~\ref{tab:strongconstant}.
\begin{table}
\tbl{Sets of the strong-coupling constants. The cutoffs $\Lambda_{\rho}$
and $\Lambda_{\omega}$ are in units of GeV.
}
{\footnotesize
\begin{tabular}{ccccccc}
\hline
&
 $g_{\rho NN}$&
 $g_{\omega NN}$&
 $\kappa_{V}$&
 $\kappa_{S}$&
 $\Lambda_{\rho}$&
 $\Lambda_{\omega}$\tabularnewline
\hline
S1 &
 2.79 &
 8.37 &
 3.70 &
 $-0.12$&
 - &
 - \tabularnewline
 S2 &
 2.79 &
 8.37 &
 6.10 &
 0 &
 - &
 - \tabularnewline
S3 &
 2.79 &
 8.37 &
 3.70 &
 $-0.12$&
 1.31 &
 1.50 \tabularnewline
 Cal &
 3.25 &
 15.58 &
 6.10 &
 0 &
 1.31 &
 1.50  \tabularnewline
\hline
\end{tabular}
\label{tab:strongconstant}
}
\end{table}

\noindent
{\bf Form factor at the strong vertex}

In Tab.~\ref{tab:strongconstant}, the set S3 introduces the cutoffs 
in the strong meson-nucleon vertices in the PNC potential. With a 
monopole-type form factor, the normal Yukawa function is modified as
\begin{eqnarray}
\frac{{\rm e}^{-m r}}{r} \rightarrow
\frac{{\rm e}^{-m r}}{r} - \frac{{\rm e}^{- \Lambda r}}{r}
\left[ 1 +
\frac{1}{2}\, \Lambda\, r \left(1 - \frac{m^2}{\Lambda^2}\right) \right].
\label{eq:monopole}
\end{eqnarray}

\noindent
{\bf Contributions with $2 \pi$ and $N^*$ intermediate states}

In order to account for the two-pion resonance nature of the $\rho$-meson,
we rely on the work presented in Ref. \refcite{cb-npb74}, based on
dispersion relations.
In this formalism, only
stable particles are involved and the $\rho$-meson appears indirectly
in the transition amplitude $N \bar{N} \rightarrow \pi \pi$ through its
propagator. To satisfy unitarity,  the width of the  $\rho$-meson has
to be accounted for. 
A background
contribution involving the exchange of the nucleon and the  $\Delta$
or $N^{*}$ resonances in the t-channel includes
the three lowest-lying resonances, 
$\Delta(1232)$, $N(1440)$ and $N(1520)$\cite{cb-npb74}.

\newpage
\noindent
{\bf Correction at the weak vertex}

As a result of a specific dynamics, the weak meson-nucleon interaction may 
acquire a momentum dependence that cannot be reduced to a monopole form 
factor. This could affect in particular the isoscalar $\rho NN$ coupling\cite{km-npa}. In momentum space, the corresponding $NN$ interaction here 
represented by the meson propagator, $(\bm{q}^2 + m^2_\rho)^{-1}$, 
in absence of form factor, can be approximately parametrized as
\begin{eqnarray}
\tilde{f}_{\rho}(\bm{q})=\left(1-2\frac{\bm{q}^{2}}{\bm{q}^{2}+\Lambda'^{2}}\right)\frac{1}{\bm{q}^{2}+m_{\rho}^{2}}.\label{eq:modp-km}\end{eqnarray}
The Fourier transformation to configuration space 
gives 
\begin{eqnarray}
\tilde{f}_{\rho+}(r)=\frac{\Lambda'^{2}+m_{\rho}^{2}}{\Lambda'^{2}-m_{\rho}^{2}}\;f_{\rho}(r)-\frac{2\Lambda'^{2}}{\Lambda'^{2}-m_{\rho}^{2}}\;f_{\Lambda'}(r).\label{eq:modr-km}\end{eqnarray}

\section{Results and discussions}

\noindent
{\bf Coupling constants and form factor at the strong vertex}

In Tab.~\ref{tab:result_coupling}, we show our results with various
strong-coupling constants and form factors.
DDH ``best-guess" values are employed for the weak couplings.
The resulting asymmetry is sensitive to the strong couplings 
as well as the form factor, but results do not fall within experimental 
error bars simultaneously.
\begin{table}[tbp]
\tbl{Sensitivity of the PNC asymmetry, $A_L (\times 10^7)$, to different
choices of strong-coupling constants or to monopole form factors,
and comparison with experiment.}
{\footnotesize
\begin{tabular}{c|cccc|c}\hline
~~Strong~~ & ~~S1~~  & ~~S2~~  & ~~S3~~  & ~~Cal~~   &  ~~Exp.~~ \\ \hline
13.6 & $-0.96$ & $-1.33$ & $-0.66$ & $-1.13$ &  $-0.95 \pm 0.15$ \\
45   & $-1.73$ & $-2.39$ & $-1.16$ & $-2.00$ &  $-1.50 \pm 0.23$ \\
221  &  0.43   &  0.75   & 0.25    &  0.52   &  $ 0.84 \pm 0.29$ \\
\hline
\end{tabular}
\label{tab:result_coupling}
}
\end{table}

\noindent
{\bf Corrections  with $2 \pi$ and $N^*$}

The effect of the $2\pi + N^*$ contribution is investigated with the
strong parameter sets S1 and S2, and DDH {}``best-guess'' values for the
weak-coupling constants.
The results are summarized in Tab.~\ref{tab:result_potential}.
In the column for $2\pi +N^*$, the numbers in the parentheses represent
the ratios $(2\pi +N^*)/({\rm bare}\, \rho)$.
The $2\pi +N^*$ contribution gives a relatively larger enhancement at 13.6 MeV
than at the remaining two energies, but as a whole, the ratios are similar.
Since the magnitudes of the asymmetry at 13.6 and 45 MeV are larger
than that at 221 MeV, similar ratios give more increase of the asymmetry
at low energies.
As a result, $2\pi +N^*$ with S1 gives asymmetries out of the
error bars at all the three energies.
For S2, the $2\pi +N^*$ contribution worsens the situation at low
energies, while keeping it at 221 MeV.
\begin{table}[tbp]
\tbl{Sensitivity of the PNC asymmetry, $A_L\, (\times 10^7)$, to the
effect of the finite $\rho$-width correction of the weak potential.}
{\footnotesize
\begin{tabular}{c|c|c|c|c|c}\hline
 & \multicolumn{2}{c|}{S1}
 & \multicolumn{2}{c|}{S2}
 &  \\ \cline{2-5}
 & bare $\rho$ & $ 2\pi + N^* $
 & bare $\rho$ & $ 2\pi + N^* $
 & Exp.
\\ \hline
13.6 & $-0.96$ &$-1.22$ (1.27)
     & $-1.33$ & $-1.70$ (1.28)
     & $-0.95 \pm 0.15$ \\
45   & $-1.73$ & $-2.14$ (1.24)
     & $-2.39$ & $-2.97$ (1.24)
     & $-1.50 \pm 0.23$\\
221  & 0.43 & 0.53 (1.23)
     & 0.75 & 0.93 (1.24)
     & $0.84 \pm 0.29$ \\ \hline
\end{tabular}
\label{tab:result_potential}
}
\end{table}

\noindent
{\bf Correction at the weak vertex}

The results with the form factor given by the chiral-soliton model 
at the isoscalar weak $\rho NN$ vertex are given 
in Tab.~\ref{tab:vertexcorrection}. The set S1 is used for the
strong parameters and DDH ``best-guess" values for the weak couplings.
The asymmetry is also sensitive to the values of the cutoff $\Lambda'$,
but in this case again, the correction at the PNC vertex does not
change the trends we have been observing in other cases.
\begin{table}[tbp]
\tbl{Sensitivity of the PNC asymmetry, $A_L\,(\times 10^7)$, to
the effect of a specific correction of the isoscalar PNC $\rho NN$-vertex.}
{\footnotesize
\begin{tabular}{c|c|c|c|c} \hline
~~~$\Lambda'$ (GeV)~~~  & ~~~bare $\rho$ ~~~ & ~~~3~~~  
& ~~~1.31~~~ & ~~~0.771~~~   \\ \hline
13.6 & $-0.96$ & $-1.04$ & $-1.33$ & $-1.69$   \\
45   & $-1.73$ & $-1.88$ & $-2.38$ & $-2.92$  \\
221  & $0.43$  & $0.47$  & $0.61$  & $0.67$
\\ \hline
\end{tabular}
\label{tab:vertexcorrection}
}
\end{table}
\section{Conclusion}
We calculated the PNC asymmetry in $\vec{p} p$ scattering at
the energies 13.6, 45 and 221 MeV.
We investigated
the role of the effects such as different strong-coupling constants,
cutoffs in the regularization of the potential,
long--range contributions to the $\rho$-exchange
PNC potential and PNC form factors of the isoscalar $\rho NN$ vertex.
The effects we considered in this work are not
helpful to solving the problem raised in the introduction. 
With this observation, we'd like to suggest the following issues:
The first one is that the value of the
$\omega NN$ coupling, its sign in particular, is correct. This implies
that present estimates are missing important contributions.
The second issue is the existence of large corrections to the
PNC single-meson exchange potential.
The last issue concerns the experiment, especially at the highest energy of 221
MeV.
Whatever the issue, they are quite interesting problems to be studied
in the future.

\end{document}